\documentclass[twocolumn,showpacs,preprintnumbers,amsmath,amssymb]{revtex4}

\usepackage{graphicx}
\usepackage{dcolumn}
\usepackage{bm}
\usepackage{amssymb}

\begin{document}


\title{From quasiperiodicity to 
high-dimensional chaos \\
without intermediate
low-dimensional chaos}

\author{Diego Paz\'o}
\email{pazo@ifca.unican.es}
\homepage{http://www.ifca.unican.es/~pazo}
\affiliation{Instituto de F\'{\i}sica de Cantabria,
IFCA (CSIC-UC), E-39005 Santander, Spain}

\author{Manuel A.\ Mat\'{\i}as}
\email{manuel.matias@ifisc.uib-csic.es}
\homepage{http://www.ifisc.uib-csic.es/}
\affiliation{IFISC, Instituto de F\'{\i}sica Interdisciplinar y Sistemas Complejos (CSIC-UIB), 
E-07122 Palma de Mallorca, Spain}

\date{\today}

\begin{abstract}
We study and characterize a direct route to high-dimensional chaos 
(i.e.~not implying an intermediate low-dimensional attractor) of a system
composed out of three coupled Lorenz oscillators. 
A geometric analysis of this medium-dimensional
dynamical system is carried out through a variety of 
numerical quantitative and qualitative techniques, that ultimately
lead to the reconstruction of the route. The main finding is that
the transition is organized by a heteroclinic explosion.
The observed scenario resembles the classical route to chaos 
via homoclinic explosion of the Lorenz model. 
\end{abstract}







\maketitle

{\bf 
Low-dimensional dissipative continuous systems, namely described 
by a d=3 phase space, may exhibit chaotic behavior, that may 
arise through a number of well characterized scenarios, including
the route from quasiperiodicity to chaos and routes involving
global bifurcations, like the route to chaos exhibited by the
Lorenz system. Things become more complicated when the dimensionality
of the phase space is increased. Interestingly, the limit in which
one has an infinite-dimensional phase space, and the system exhibits
the so-called spatio-temporal chaotic behavior, is well characterized
through the ergodic theory of dynamical systems, through quantifiers
like dimension, entropy, and Lyapunov exponents (a Lyapunov density
in continuous systems). Less explored is the case in which one has a
phase space that has an intermediate dimension between these two
limits. In this case, one may have high-dimensional chaotic behavior,
and an obvious possible route is a cascade in which the dimensionality
of the high-dimensional attractor increases sequentially. Still, other
possibilities may take place, like the scenario discussed in the present
work, in which quasiperiodic attractors participate, though in a manner
different to the {\it classical\/} quasiperiodic route to chaos. Instead,
a high-dimensional chaotic attractor is created in a global bifurcation.
Interestingly, the route exhibits a regime in which a three-dimensional
quasiperiodic attractor is stable, in apparent contradiction to {\it
common wisdom\/} interpretation of Ruelle-Takens-Newhouse Theorem.
}

\section{Introduction}

Nonlinear dynamics was considerably boosted 
by the recognition of chaos as an ubiquitous phenomenon out of the 
linear realm. Chaos theory succeeded to 
identify the universal routes through which regular motion
may become chaotic. In
dissipative dynamical systems, chaos
appears through a few well characterized routes (or scenarios)
\cite{Eckmann81} like the following: (a) the period-doubling cascade
\cite{Feigenbaum78}; (b) the intermittency route \cite{Pomeau80};
(c) the route involving quasiperiodic tori
\cite{Newhouse78,curryorke}; (d) the crisis route
\cite{grebogi3D}; (see also \cite{berge,Ott} for a survey).
Another possibility is that chaos sets in through a global
connection to a fixed point, as is the case \cite{Sparrow} of the Lorenz 
system, and of Shilnikov chaos \cite{AAIS,nato}.

More recently, some studies have been devoted to the study of
high-dimensional chaos, that here we shall define as chaotic
motion inside an attractor of typical dimension $d>3$ (as a typical one we may  
take the information dimension $D_1$).
An obvious possibility to transit to high-dimensional chaos
is to take as starting point a low-dimensional chaotic attractor. 
This possibility has been often considered in the context of
desynchronization between coupled chaotic oscillators. A possible
scenario, and probably the most studied one
\cite{Harrison99,Kapitaniak00} involves generating a hyperchaotic
attractor---an attractor with two or more
positive Lyapunov Exponents (LEs)---from a
low-dimensional chaotic attractor. 
Another class of high-dimensional
chaotic attractors exhibits two (at least very approximately) null Lyapunov exponents.
As shown in
\cite{matiasepl97,sanchezIEEE00,Yang00} one may encounter this situation in rings 
of asymmetrically (e.g.~unidirectionally) coupled chaotic systems in which
a symmetric Hopf bifurcation renders unstable the synchronized state
\cite{matiasprl97,matiasprl98}. 

Less obvious is the possibility of a direct transition to
high-dimensional chaos without an intermediate low-dimensional
chaotic attractor. Restricting to autonomous ordinary differential
equations, the transition to high-dimensional chaos has been found
to be associated to quasiperiodicity. Thus, the works by Feudel
{\it et al.}~\cite{feudel93} and Yang~\cite{Yang00} report a
transition from two- and/or three-frequency quasiperiodicity 
to high-dimensional chaos. 
A more geometrical view is in the work by Moon \cite{moon97}, that 
describes a route (to high-dimensional
chaos) from a two-dimensional torus,
through a global bifurcation that comprises a double homoclinic connection
to a limit cycle
(this structure
amounts to adding one dimension to each building block of the homoclinic
route to chaos in the Lorenz model). 
Recently~\cite{epl}, we reported a route to chaos whose
most novel aspect is that it implies the sudden
creation (i.e. with no mediating low-dimensional chaos) of a
high-dimensional chaotic attractor with $D_1>4$. In summary, in this route
first a high-dimensional chaotic (nonattracting) set is created in a global
bifurcation. Then, this set is rendered stable, becoming a high-dimensional
chaotic attractor \footnote{that coexists with two already existing symmetry 
related nonchaotic attractors}, in a boundary crisis.

The purpose of this paper is to explain in more detail our findings
in Ref.~\cite{epl}. In Sec.~\ref{systoverall}
the system is described and an overall picture of the route to
high-dimensional chaos is presented. Then, 
Sec.~\ref{quasip} discusses at some
detail the appearance and the robustness of the two- and
three-frequency quasiperiodic attractors found in the system.
Later, Sec.~\ref{numevid} presents the numerical evidences that
have been used to understand the complex route to chaotic behavior
presented by the system. Section~\ref{return_map} presents a
characterization of the route to high-dimensional chaos through a
return map similar to that of the Lorenz system. The goal of
Section \ref{chaosroute} is to present and discuss at
some length the route to chaos exhibited by the system. Finally,
Secs.~\ref{further} and \ref{discussion} present the final remarks
on this work and the conclusions, respectively.

\section{System and overall picture} \label{systoverall}

The system studied in the present work is a $9$-dimensional dynamical 
system considered here is formed by three Lorenz \cite{lorenz63}
oscillators coupled according to the partial replacement coupling method 
of Ref.~\cite{guemez95}.
The oscillators are unidirectionally coupled in a ring geometry such that,
\begin{eqnarray}
\left.
\begin{array}{rcl}
\dot{x_j}&=&\sigma(y_j-x_j)\\
\dot{y_j}&=&R\,\underline{x_j}-y_j-x_j\,z_j\\
\dot{z_j}&=&x_j\,y_j-b\,\,z_j
\end{array}
\right\} \quad j=1,\ldots,N=3 \label{eqlor}\ ,
\end{eqnarray}
where $\underline{x_j}=x_{j-1}$ for $j\neq 1$, define the
coupling, and $\underline{x_1}=x_3$ the ring geometry of the
system. In our study two of the parameters are fixed
as in \cite{pazoijbc01,epl}): $\sigma=20$, $b=3$, while $29<R<40$.
The study of this system has been suggested by the results of
the experimental study of rotating waves for three coupled Lorenz oscillators
corresponding to these parameters \cite{sanchezijbc99,sanchezetal}.

A useful representation in the study and
characterization of discrete rotating waves are the
(discrete) Fourier spatial modes
\cite{heagy94,matiasprl97}. These modes are defined as,
\begin{equation}
{\bf X}_k={1\over N} \sum_{j=1}^{N} {\bf x}_j \exp\left[{{2\pi i
(j-1) k}\over N}\right]\ , \label{modes}
\end{equation}
where $N=3$ (as already indicated) and $i$ is the imaginary unit.
In terms of these modes (${\bf X}_0 \in \mathbb{R}^3,{\bf X}_1 \in \mathbb{C}^3$), 
the evolution equations are,
\begin{eqnarray}
\left.
\begin{array}{rcl}
\dot{X_0}&=&\sigma(Y_0-X_0)\\
\dot{Y_0}&=&R\,X_0-Y_0-X_0\,Z_0-X_1\,Z_1^*-X_1^*\,Z_1\\
\dot{Z_0}&=&X_0\,Y_0-b\,\,Z_0+X_1\,Y_1^*+X_1^*\,Y_1\\
\dot{X_1}&=&\sigma(Y_1-X_1)\\
\dot{Y_1}&=&{\tilde R}\,X_1-Y_1-X_0\,Z_1-X_1\,Z_0-X_1^*\,Z_1^*\\
\dot{Z_1}&=&X_0\,Y_1+X_1\,Y_0-b\,\,Z_1+X_1^*\,Y_1^*
\end{array}
\right\} \label{eqmod}\
\end{eqnarray}
with ${\tilde R}=R\exp(2\pi i /3)$, and where $X_1^*$ denotes the complex
conjugate of $X_1$.

The qualitative description of the behaviors exhibited by the
system in this range of parameters is as follows.
The three Lorenz oscillators exhibit chaotic synchronization
for $R< R_{sc}\approx 32.82$. At $R_{sc}$ the system exhibits 
a Hopf bifurcation directly
from a chaotic state, yielding a behavior, that was first found in rings of
unidirectionally coupled Chua's oscillators and called a Chaotic
Rotating Wave (CRW) \cite{matiasepl97}. The Hopf bifurcation
exhibited by the system is called symmetric \cite{GolStewARMS85,CollinsStewart}, 
as it originates from the cyclic ($\mathbf{Z}_3$) symmetry of the ring.
Close to the bifurcation the CRW is the combination
of the chaotic dynamics in a Lorenz attractor and 
a superimposed oscillation created by the Hopf bifurcation, that
occurs in the subspace transverse to the synchronization manifold
($k=1$ mode).
The oscillation created in the symmetric Hopf bifurcation is
characterized by 
a phase difference of $2\pi/3$ ($2\pi/N$ in the general
$N$-oscillator case) between neighbor oscillators, and 
this behavior is
the discrete analog of a traveling rotating wave \cite{POM}. This
picture is valid close to onset, $R \gtrsim R_{sc}$, and when
increasing $R$ one observes that this behavior changes. For
$R>35.26$ the behavior of the system becomes periodic, with a
waveform characteristic of the $k=1$ Fourier mode.
Actually, two 
mirror limit-cycle attractors, called Periodic Rotating Wave (PRW) in
Ref.~\cite{matiasprl97,sanchezpre98}, are found
in the range $R\in [35.26,39.25]$ (approximately).
At $R=R_{pitch} \approx 39.25$ both solutions merge giving rise,
through a pitchfork bifurcation, to a centered stable symmetric
periodic behavior.

\begin{figure}
\includegraphics[width=0.9\columnwidth]{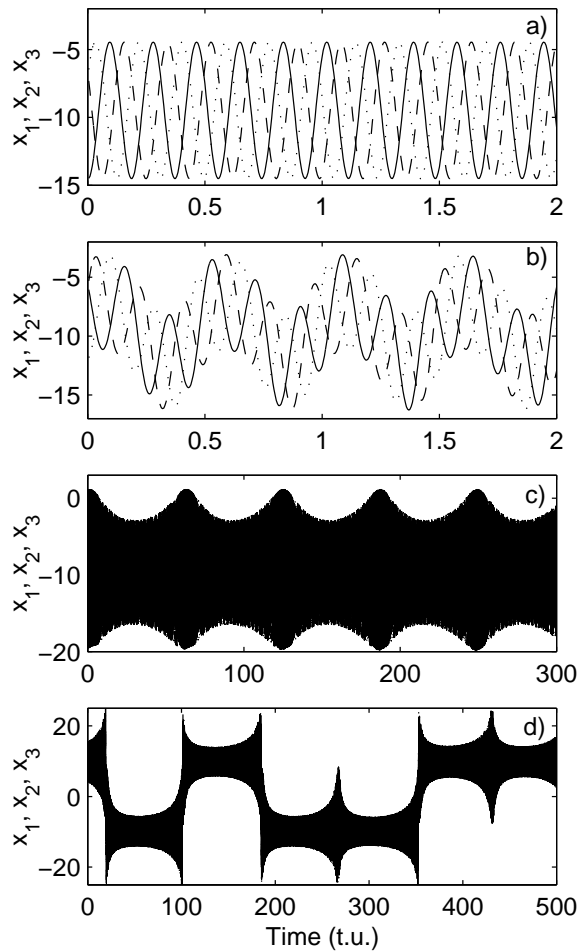}
\caption[]{Time series of the system Eq.~(\ref{eqlor}): 
(a) $R=35.5$, Periodic Rotating Wave; (b)
$R=35.2$,  $\mathbb{T}^2$; (b) $R=35.095$,  $\mathbb{T}^3$; (d)
$R=35.093$, Chaotic Rotating Wave. Note the different time scale
for each panel.} \label{fig1osc}
\end{figure}

We shall not describe here the transition from synchronous chaos
to CRW; instead, we shall focus on the transitions from PRW to CRW, 
i.e.~we go `from order to chaos' decreasing $R$. The reader should notice that by CRW
we refer to a high-dimensional chaotic attractor characterized by
a oscillation with a $2\pi/3$ phase shift between neighboring
oscillators superimposed to an underlying chaotic behavior. 
The temporal series for the different behaviors
studied in this work are shown in Fig.~\ref{fig1osc}. Note that
the third frequency, that appears as the $\mathbb{T}^3$ is born,
manifests as a very slow modulation on the size of the former
$\mathbb{T}^2$ (the time scale has been broadened in
Fig.~\ref{fig1osc}(c) in order to allow the observation of this slow 
scale).

\begin{figure}
\includegraphics[width=0.9\columnwidth]{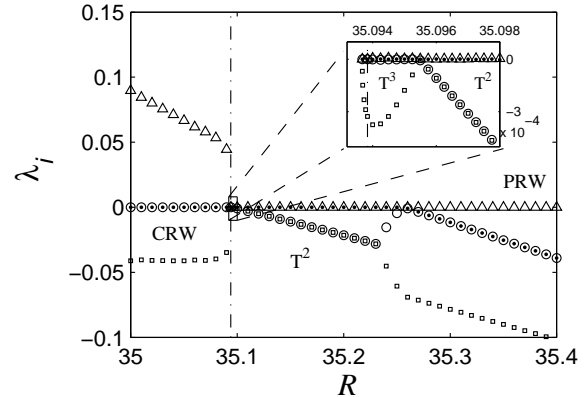}
\caption[]{The four largest Lyapunov exponents
($\lambda_{i=1,...,4}$) as a function of the parameter $R$. Four
regions can be distinguished according to their different Lyapunov
spectra. Note that for the high-dimensional chaos we obtain
$\lambda_2=\lambda_3=0$ and $\lambda_1 \gtrsim |\lambda_4|$ which
implies, according to the Kaplan-Yorke conjecture, an information
dimension $D_1\gtrsim4$. The dash-dotted line indicates the value of $R$
where the chaotic {\em attractor} is born.} \label{lyap_tran}
\end{figure}

A view of the parameter region that will be
considered in this work can be found in Fig.~\ref{lyap_tran}, in
which the Lyapunov spectrum corresponding to the four largest
Lyapunov exponents is presented. From right to left, the two
symmetry-related limit-cycle attractors exhibit a
(supercritical) Hopf bifurcation when $R$ is decreased, namely at
$R=R_{h1}\approx35.26$, yielding two symmetry-related
two-frequency quasiperiodic attractors (with two null LEs). 
Lowering $R$ further the system exhibits another
(supercritical) Hopf bifurcation, at $R=R_{h2}\approx35.0955$,
that yields two mirror three-frequency quasiperiodic
attractors (three vanishing LEs). Lowering $R$ furthermore the
system exhibits a boundary crisis, at $R=R_{bc}\approx35.09384$,
in which the chaotic attractor is born (or destroyed seen from
the opposite side). The chaotic attractor is characterized by two
vanishing (at least, very approximately) Lyapunov exponents, and a
single positive LE that is larger than the absolute value of the
fourth LE. This implies, according to the Kaplan-Yorke conjecture\footnote{$D_1 = 
K+\sum_{j=1}^K (\lambda_j/\vert
\lambda_{K+1}\vert)$, being $K$ the largest integer such that
$\sum_{j=1}^K \lambda_j\ge 0$, where the $\lambda_j$ are the
Lyapunov exponents ordered from larger to smaller.},
that the information dimension $D_1 > 4$. As we shall see below (see
Sec.~\ref{branched_manifold}) this dimension is genuine. 

The presence of 2-D and 3-D quasiperiodic attractors may lead to
think that chaos appears through a quasiperiodicity transition
to chaos (see Sec.~\ref{quasip}). However, it will be shown
below that the chaotic attractor is created at a boundary crisis,
and coexists with the two $\mathbb{T}^3$ attractors until the
latter are destroyed as each of them collides with a twin unstable
$\mathbb{T}^3$, at $R=R_{sn}=35.09367$. We shall show (from
Sec.~\ref{numevid}) that, indeed, the system exhibits a global
bifurcation ---analogous to what happens for the Lorenz
system \cite{Sparrow}---
that implies, in first approximation, 
the sudden creation of an infinite number
of unstable 3-D tori. 

A schematic diagram of the whole set of bifurcations linking the
PRW and synchronous chaos is shown in Fig.~\ref{diagram1}. As
mentioned above, in the interval of $R$ where the CRW is found,
the shape of the attractor changes. Anyway, in this paper we are
not interested in the transitions between different types of
chaotic rotating wave (see \cite{sanchezetal} for such a study).

\begin{widetext} 
\begin{center}
\begin{figure}
\includegraphics[width=0.9\columnwidth]{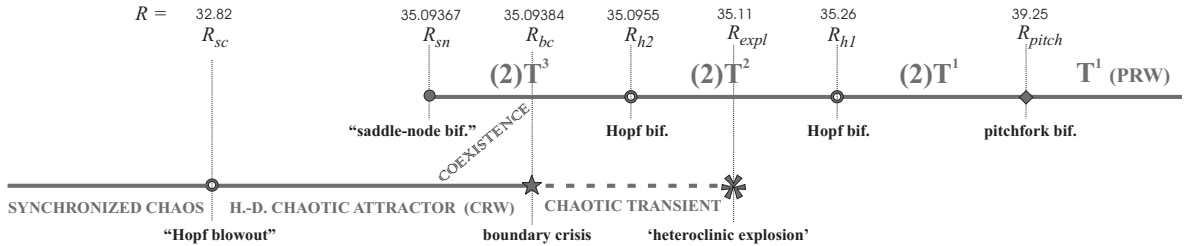}
\caption[]{Diagram representing schematically the transitions from
synchronous chaos (left) to a PRW (right).} 
\label{diagram1}
\end{figure}
\end{center}
\end{widetext} 

\section{Transition to Quasiperiodic behavior} \label{quasip}

As described above, the (symmetric) periodic rotating wave, that
is the starting point of our analysis, is replaced after a sequence
of bifurcations by two mirror three-frequency quasiperiodic attractors,
$\mathbb{T}^3$. The involved transitions are a pitchfork
bifurcation, in which two (asymmetric) periodic rotating waves are
born, and then two consecutive Hopf bifurcations. 
A small note is in place about the stability of this
type of unusual attractors as it
has been stated that $\mathbb{T}^3$ attractors are intrinsically fragile
according to  Newhouse-Ruelle-Takens (NRT) Theorem \cite{Newhouse78}.

First of all, one can present numerical evidence to show that,
indeed, three-frequency quasiperiodic attractors are observed
robustly in a finite parameter range. 
In particular the inset of 
Fig.~\ref{lyap_tran} shows results corresponding to the calculation
of the Lyapunov spectrum in the $\mathbb{T}^3$ parameter range. As there are
long chaotic transients in this region,
the calculations were performed using long transients
(of the order of $10^6$ t.u.), while another $10^6$ t.u. were used
to calculate the spectrum. 

On the other hand, and as already reported in
Ref.~\cite{pazoijbc01}, it can be numerically checked that the
system exhibits three-frequency quasiperiodic behavior by looking
at the Poincar\'e cross section of a $\mathbb{T}^3$ torus in phase space.
Namely, in Fig.~2 of Ref.~\cite{pazoijbc01} the Poincar\'e sections
of a $\mathbb{T}^2$ and of a $\mathbb{T}^3$ torus, respectively, are 
presented. While the former is a circle, the latter is clearly a 
two-dimensional, $\mathbb{T}^2$, torus.
In the literature Ref.~\cite{Yang00} reports the existence of three-frequency
quasiperiodicity in a ring of coupled Lorenz oscillators (in a different
parameter regime), while a case of a third very low frequency associated 
to a $\mathbb{T}^3$ torus has been observed in 
Refs.~\cite{wang,LopezMarques00,LopezMarques01}.

In trying to justify the robust existence of $\mathbb{T}^3$ dynamics
in the light of the NRT Theorem, a first observation is that
in our simulations we have not detected resonances in
the $\mathbb{T}^2$ regime. This may occur due to the fact that
along $R$, the winding number does not crosses hard resonances,
the smallest denominator is $q=11$ (corresponding to a winding
number of $4/11$), so the resonance horns may be quite small. But 
also it is important to notice that 
the absence of resonances in rotating waves is a characteristic
feature of systems with rotational, $SO(2)$, symmetry~\cite{rand}. 
For instance
in a homogeneous excitable medium, the compound rotation of a spiral wave 
lacks of frequency lockings due to this symmetry~\cite{barkley95}. Therefore
the stability of the $\mathbb{T}^3$ attractors would be subsequent to the 
robustness of the quasiperiodic dynamics on the two-tori. In a sense
our three-frequency quasiperiodic attractors would be as robust
as two-frequency quasiperiodic attractors. 
It is clear that in our system this rotational symmetry is only approximate,
but one may consider the discreteness effect as a small perturbation. 
Also the third frequency is very
small what precludes strong resonances with small denominators:
thus, for $R=35.095$ [Fig.~\ref{fig1osc}(c)] we have $f_1=5.3\ldots$,
$f_2= 1.87\ldots$, and $f_3= 0.016\ldots$.
These arguments support the existence of a 
range of three-frequency quasiperiodic behavior.

\section{Numerical evidences of the route to chaos exhibited by the system}
\label{numevid}

In this section we are going to analyze the route through which a
chaotic attractor is born in this system. 
First of all, 
we shall reduce the dimensionality of the problem by eliminating the fast
frequency, namely the one involved in the phase shift by $2\pi/3$ in neighbor
oscillators, as it leads to
a conserved quantity (this time lag) and consequently a vanishing
(or almost vanishing) Lyapunov exponent. In the mode representation of
Eq.~(\ref{eqmod}) this amounts to perform a cut through the Poincar\'e section $Im
(X_1)=0, Im({\dot X}_1)>0$. So, in visualizing objects cycles will become fixed
points, $\mathbb{T}^2$-tori cycles and $\mathbb{T}^3$-tori will
become $\mathbb{T}^2$-tori, although we shall refer to these
objects corresponding to the complete phase space, and not to the result
of the dimensionality reduction achieved via sectioning.

\subsection{Coexistence between 3D-torus and CRW}
\label{coex}

The first important remark about our system of three coupled
Lorenz oscillators is that the two $\mathbb{T}^3$ attractors are
not directly involved in the birth of the high-dimensional chaotic
attractor. Continuation of the $\mathbb{T}^3$ attractors shows
that they exist above $R_{sn}=35.09367$.
This implies that the system exhibits
multistability between the high-dimensional chaotic and the
$\mathbb{T}^3$ attractors in the range $R_{sn}<R<R_{bc}$, with the
latter attractors having a small basin of attraction.
This remark is important, because it implies that the
high-dimensional chaotic attractor is not created through some
route involving the $\mathbb{T}^3$ attractors.

\begin{figure}
\includegraphics[width=0.9\columnwidth]{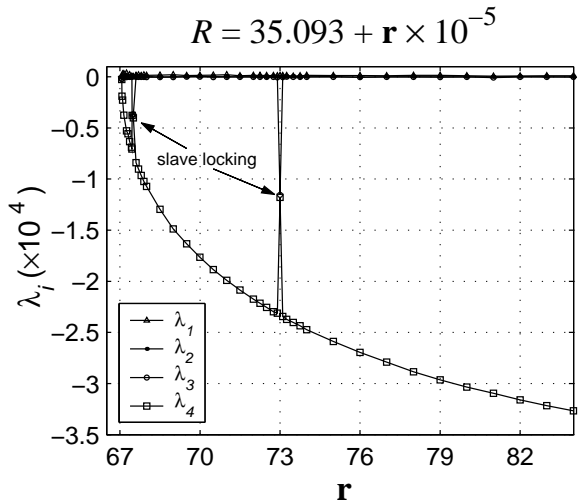}
\caption[]{Blowout of the four largest Lyapunov exponents for the
$\mathbb{T}^3$ attractors in the region in which they coexist with
the chaotic attractor. The fourth Lyapunov exponent approaches
zero in a square-root fashion, as expected for a saddle-node bifurcation.} 
\label{lyapunovd}
\end{figure}

A detailed view of the behavior of the four largest Lyapunov
exponents for the $\mathbb{T}^3$ attractors is represented in 
Fig.~\ref{lyapunovd}. It can be seen
that close to $R_{sn}$ the fourth Lyapunov exponent
exhibits a square-root profile as expected for a saddle-node
bifurcation. This suggests that each $\mathbb{T}^3$ attractor is
approaching an unstable $\mathbb{T}^3$. Also one can appreciate
quite clear lockings\footnote{Recall that the classical theory
for two-frequency quasiperiodic attractors tells us that when a two-torus
locks, a pair of stable-saddle orbits are born on its surface
through a saddle-node  bifurcation. According to this, one of the
Lyapunov exponents becomes slightly negative, indicating the small
attraction along the torus surface to the stable limit cycle (the
new attractor). Generically, when a parameter
varies, the torus visits some (formally infinity) Arnold tongues
where its rotation number is a rational number; accordingly 
a stable periodic orbit appears on its surface. Analogous
resonances appear for 3D-tori \cite{Baesens}.} where the third and fourth
Lyapunov exponents become equal.

\begin{figure}
\includegraphics[width=0.9\columnwidth]{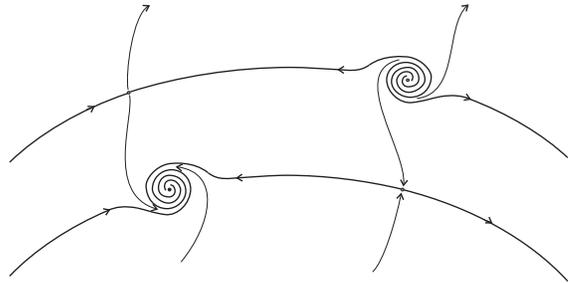}
\caption[]{Schematic of a `slave locking'. A portion of a section
of the tori (stable and unstable) are shown with bold lines. Tori
are not differentiable at the foci.} \label{slave_locking}
\end{figure}

As $R$ approaches $R_{sn}$ the existence of a smooth invariant
3D-torus cannot be guaranteed, because the
strength of the normal contraction can be of the same order than
the rate of attraction of the 2D-torus arising in each locking \cite{ChowHale}. 
Thus, in Fig.~\ref{slave_locking} a possible scenario for the interaction
of the tori, that would explain the lockings observed in
Fig.~\ref{lyapunovd} is presented. Because of the extremely small
transversal stability of the $\mathbb{T}^3$ (represented by the
fourth exponent), for $R \gtrsim R_{sn}$, we conjecture that 
the transversal
direction `slaves' the tangential one (represented by the third
exponent). Then, we observe two identical non-vanishing exponents
that may indicate the existence of a stable focus-type $\mathbb{T}^2$
on the surface of the $\mathbb{T}^3$. The $\mathbb{T}^3$ continues
to exist but is non-differentiable at the stable $\mathbb{T}^2$
located on its (hyper)surface.

The participation of the unstable 3D-torus in the final
annihilation of the stable one is not trivial (the existence of
the unstable $\mathbb{T}^3$ is supported by another numerical evidence,
discussed in Sec.~\ref{bcrisis}), but we want to recall that 
saddle-node bifurcations of $\mathbb{T}^2$ tori have been reported 
\cite{Ashwin98} (called bubbles in this reference), and due to the 
special character of the $\mathbb{T}^3$ tori that
we have here (with one oscillation that, in some sense, is orthogonal to
the others and also almost conserved, cf. Sec. \ref{quasip}), and given 
the numerical evidence reported here, we think that we have found a
saddle-node bifurcation of $\mathbb{T}^3$ tori.


\subsection{Route to chaos in the Lorenz model 
through homoclinic explosion and boundary crisis}

As the route studied in this work presents some
similarities with the classical route to chaos for an isolated
Lorenz system, we are going to draw some useful analogies with
that system, although it is warned that the analogy is not
complete (otherwise, we would not have a new route to chaos). 
As it can be found in textbooks~\cite{Ott}, the Lorenz system
exhibits several routes to chaos.
The classical one 
is the route to chaos 
through a double homoclinic connection
of the saddle point at the origin. Fixing $\sigma$ and $b$,
the double homoclinic orbit occurs at a particular value 
$R=R_{HOM}$\footnote{for
$\sigma=10$ and $b=8/3$: $R_{HOM}\approx 13.926$, $R_{BC}\approx
24.06$, and $R_{H}\approx 24.74$ \cite{Sparrow}.}.
It has the shape of the butterfly (taking into account how the
unstable directions reenter the saddle point at the origin, being
this dictated by the reflection symmetry of the system), and this gives birth
\cite{AAIS} to a chaotic set for $R>R_{HOM}$ (a
homoclinic explosion \cite{yorkey,Sparrow}). The closure of this
set is formed by the infinite number of unstable periodic orbits (UPOs)
that can be classified according to their symbolic sequences of
turns around the right ($\mathbf{R}$) and the left ($\mathbf{L}$)
fixed points ($C_+$ and $C_-$)~\cite{Sparrow}. The appearance of
these UPOs reflects the dramatic change undergone by the stable
manifold of the fixed point that allows initial conditions at one
side of the phase space jump to the other side (before falling to
one of the two symmetry related stable fixed points $C_+$ and
$C_-$), being these jumps impossible for $R<R_{HOM}$. For a larger
value of $R$, $R=R_{BC}$, the chaotic set becomes stable in a
boundary crisis, that occurs precisely when the two shortest
symmetry-related length-$1$ unstable periodic orbits ($\mathbf{R}$
and $\mathbf{L}$) located at each lobe, and that at $R=R_{HOM}$
coincide with the two homoclinic orbits, shrink, such that the
chaotic set has a tangency with these two orbits~\cite{Sparrow}.
At this value of $R$ there exists a double heteroclinic connection
between the equilibrium at the origin and the mentioned length-$1$
UPOs. For $R_{BC}<R<R_{H}$ these two orbits (and their respective
tubular stable manifolds) form the basin boundaries between the
chaotic attractor and the two stable asymmetric fixed points
$(C_{\pm})$, and, consequently, the system exhibits
multistability.  These two fixed points loose their stability in a
a subcritical Hopf bifurcation, that occurs when 
the two mentioned length-$1$ orbits shrink to a point
coinciding with $(C_+,C_-)$.

\subsection{Heteroclinic explosion}

In our system one can find a value of
$R=R_{expl}\approx35.11$ that defines a clearcut transition in the
way in which transients approach the attractors (that for $R \sim
R_{expl}$ are two $\mathbb{T}^2$-tori). For $R>R_{expl}$ the basin
of attraction of each $\mathbb{T}^2$ is quite simple. But below
$R_{expl}$ trajectories may tend asymptotically to one of the
$\mathbb{T}^2$ after visiting the neighborhood of the other torus.

\begin{figure}
\includegraphics[width=0.9\columnwidth]{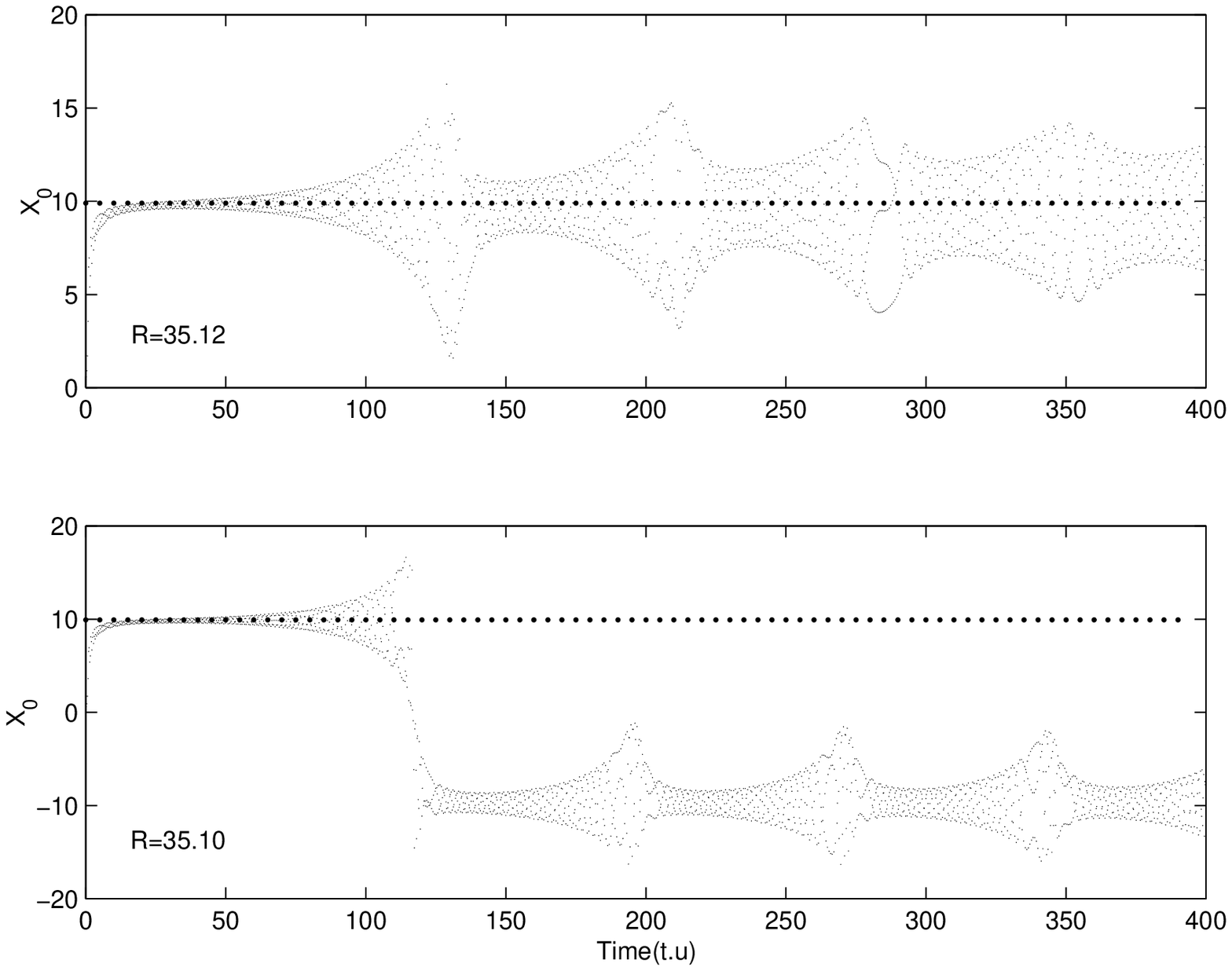}
\caption[]{Numerical experiment showing the time series of the coordinate $X_0$
at the Poincare section for 
trajectories starting in
an initial condition in the symmetric unstable PRW: (a) for
$R=35.12$, a condition before the explosion (b) for $R=35.10$ just
past the explosion. The $X_0$ coordinate of one of the two asymmetric PRWs is 
denoted with a bold dotted line.} 
\label{explosion}
\end{figure}

Following the analogy with the Lorenz system, we conjecture that
precisely at $R=R_{expl}$ a global bifurcation occurs, and past
this value ($R<R_{expl}$) an infinite number of unstable objects
are created. To check this we stabilized the symmetric PRW  (an
unstable fixed point in the Poincar\'e map) by a Newton-Raphson
method and observed the fate of the trajectories starting from
(approximately) that solution. In Fig.~\ref{explosion} the
evolution of the trajectories for two values of $R$ above and
below $R_{expl}$ is shown. After approaching one the asymmetric
PRWs (its location is marked with
a bold dotted line) the trajectory
jumps or not to the other side. The result obtained for $R_{expl}$
is the same if one takes as starting point one of the asymmetric
PRWs.

\subsection{Four-dimensional branched manifold}
\label{branched_manifold}

The third important numerical remark refers to the Lyapunov
spectrum (Table I) of the different attractors in the
multistability region (of course, and due to symmetry, the
Lyapunov spectra of the two $\mathbb{T}^3$ attractors are
identical). As one can see, fifth to ninth LEs are quite similar,
indicating that both attractor types are approximately embedded in the same
four-dimensional space, a subset of the total phase space.

\begin{center}\label{tabla}
\begin{tabular}{|c|c|c|} \hline
  $\lambda_i$ & CA & $\mathbb{T}^3$ \\ \hline
  $\quad\lambda_5\quad$ & $\quad -5.254(7) \quad$& $\quad -5.202(7) \quad$ \\
  $\lambda_6$ & $\quad-5.254(8) \quad$ & $\quad-5.202(7) \quad$  \\
  $\lambda_7$ & $\quad-18.612(4) \quad$ & $\quad-18.651(9) \quad$\\
  $\lambda_8$ & $\quad-18.612(4) \quad$ & $\quad-18.651(9) \quad$\\
  $\lambda_9$ & $\quad-24.273(4) \quad$ & $\quad-24.290(4) \quad$\\ \hline
\end{tabular}

TABLE I: The five smallest Lyapunov exponents for the two
attractors (CA: chaotic, and $\mathbb{T}^3$: three-frequency
quasiperiodic) coexisting at $R=35.0938$.
\end{center}

Thus, we postulate that the dynamics can be simplified to the study of
a four-dimensional branched manifold. A theoretical justification
of this statement is impossible, but one can argue along the lines
of (a higher-dimensional generalization of) the Birman-Williams
Theorem \cite{gilmore,BirmanWilliams1} (this theorem has proven to be 
very useful, even if it requires the strange attractors to be (uniformly) 
hyperbolic what is usually not fulfilled; notice, however, some recent rigorous 
advances \cite{singhyp} on the theory of singular hyperbolic attractors, 
that encompass the Lorenz equations \cite{lorenz63}.

In our case we can
argue that the high-dimensional chaotic attractor has, according
to Kaplan-Yorke conjecture \cite{kapyor}, a value of the
information dimension $D_1\gtrsim 4$ (cf.~Fig.~\ref{lyap_tran}). 
We have also measured the correlation dimension $D_2$ 
(through Grassberger-Procaccia algorithm)
obtaining also a value close to $4$, namely $D_2=3.96\pm0.05$),
(recall that $D_1\geq D_2$).

A more correct picture of the attractor amounts to consider that
this $4$-dimensional manifold is actually composed of many
{\it thin\/} leaves. Of course this
reduced $4$-dimensional picture can be only considered to be a
more or less faithful representation of the system. 
In the epochs in which a trajectory jumps to the other
side (subspace), which means reinjections through an extra
dimension, it is when the existence of branching is needed in
order to the trajectory not to intersect itself. In our case the
`tear point' is the symmetric PRW. This is in complete analogy
with what happens with the Lorenz system, where the attractor can
be understood as a template composed of a branched two-dimensional
manifold with a tear point at the origin (cf. Fig.~20(b) in 
\cite{gilmore}). Rotations around one of
the lobes are roughly planar, but reinjections between the two
(planar) lobes, that form themselves an angle, involve the third
dimension.

An important remark concerning this $4$-dimensional picture is
that, in this space, a $\mathbb{T}^3$-torus divides
a $4$-dimensional manifold 
in two regions (just the same as a cycle divides
a surface). Then in the regime with coexistence, between
chaos and three-frequency quasiperiodicity, it is the pair of
(conjectured) unstable 3D-tori what defines the basin boundary of the chaotic
attractor. In the Lorenz system the length-$1$ unstable periodic
orbits divide the $2D$-manifold in three regions; and also they 
acts as the basin boundaries, when the chaotic and the fixed point 
attractors coexist. 
In the transient chaos region,
trajectories spiral in one lobe away from the fixed point (say
$C_+$) because of the repulsive effect of the UPO surrounding that
fixed point. After some turns the trajectory jumps to the other
lobe by using the third dimension (i.e. by virtue of the branching).
If the trajectory approaches close enough to $C_-$ surpassing the
`barrier' constituted by the length-$1$ UPO, it is captured by
this fixed point and no more jumps occur (i.e. the chaotic
transient finishes). Analogously, the unstable
$\mathbb{T}^3$-torus acts as a dividing hypersurface for
trajectories in the $4$-dimensional branched manifold, as
discussed above.

\subsection{Boundary crisis and power law of chaotic transients}
\label{bcrisis}

\begin{figure}
\includegraphics[width=0.9\columnwidth]{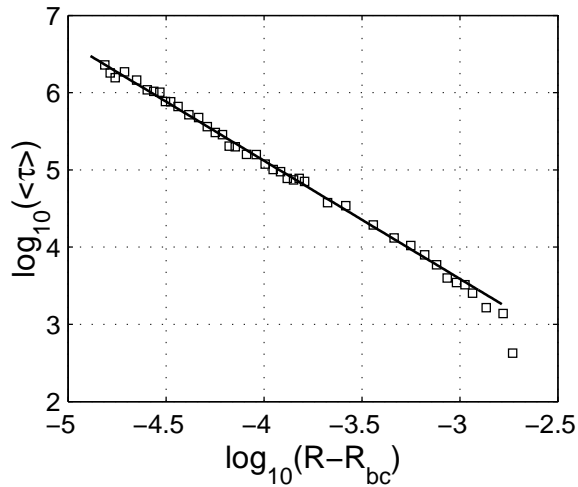}
\caption[]{Log-log representation of the average chaotic transient
as a function of the distance to the critical point $R_{bc}$. Each
point is an average over 100 realizations. Data fit to a straight
line of slope $\gamma=-1.53\pm0.06$.}\label{scaling}
\end{figure}

 The fourth remark regards numerical studies for $R
\gtrsim R_{bc}$. We have measured the average time of the chaotic
transients ($\left< \tau\right>  $) when approaching $R=R_{bc}$.
We observe that these transients diverge satisfying a power law 
($\left<\tau\right>  \propto (R-R_{bc})^\gamma$), with 
$\gamma=-1.53\pm0.06$, for an asymptotic value $R=R_{bc}=35.093838$,
as expected for a boundary crisis~\cite{grebogi87}, see 
Fig.~\ref{scaling}\footnote{A quantitative explanation of the scaling
exponent $\gamma$ along the lines of Ref.~\cite{grebogi87} has
been attempted, but the high-dimensional nature of the system casts
doubts on its reliability.}.

\begin{figure}
\includegraphics[width=0.9\columnwidth]{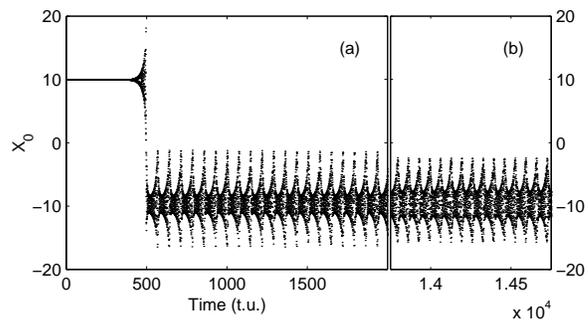}
\caption[]{Numerical experiment showing a trajectory (only points
at the Poincar\'e section are shown) starting in an initial
condition at one of the asymmetric unstable PRWs for $R=35.09389
\gtrsim R_{bc}$. The unstable 3D-torus, larger, is seen initially,
but finally (part (b)) the system decays to the smaller stable
3D-torus. Note that the minimum width along time is larger for the
stable 3D-torus as expected for a torus located inside another
(see the discussion about the branched manifold).}
\label{boundary_crisis}
\end{figure}

Inspired in the boundary crisis occurring in the Lorenz
system~\cite{kaplan79lorenz,grebogi87,Sparrow}, we took one initial
condition in one of the asymmetric (already unstable) Periodic
Rotating Waves for different values of $R$. When $R$ is
above $R_{bc}$ (but below $R_{expl}$),
the system goes away of the limit cycle (due to numerical roundoff errors), and 
jumps to the $\mathbb{T}^{2}$ (or $\mathbb{T}^{3}$) attractor 
located at the other side of phase space. 
Contrary to what happens to typical initial conditions for those
$R$ values, the chaotic transient is {\em not} observed, as may be seen 
in Fig.~\ref{boundary_crisis} (the PRW is a fixed point in the figure, 
as it has been stroboscopically cut through the Poincar\'e hyperplane
$Im(X_1)=0$). 
For $R$ just above $R_{bc}$ (cf.~Fig.~\ref{boundary_crisis})
the trajectory seems to approach a $\mathbb{T}^3$
before `falling' to the $\mathbb{T}^3$ attractor. Thus, we conjecture
that is the unstable $\mathbb{T}^3$ which constitutes the basin
boundary of the chaotic attractor and the object involved in a
global connection that marks the birth (or the death, depending on
the viewpoint) of the chaotic attractor. Notice also, that the
stable and the unstable  $\mathbb{T}^3$-tori have quite similar
sizes. Ultimately, at $R_{sn}$, the multistability region has an
end when the two $\mathbb{T}^3$-tori annihilate in a saddle-node
bifurcation (actually there are two mirror bifurcations, 
as previously discussed in Sec.~\ref{coex}).

\section{Description in terms of a return map}
\label{return_map}

Lorenz \cite{lorenz63} described a nice technique for reducing the complexity 
of the solutions of the Lorenz equations. By recording the successive
peaks of the variable $z(t)$, he reduced the dynamics of the
Lorenz system to a one dimensional map. Denoting the $n$th
maximum of $z(t)$ by $M_n$, he plotted successive pairs
($M_n,M_{n+1}$), finding that points lay (very
approximately) along a $\Lambda$-shaped curve. In this way, the
dynamics is reduced to the ``Lorenz map": $M_{n+1}=\Lambda(M_n)$.

\begin{figure}
\includegraphics[width=0.9\columnwidth]{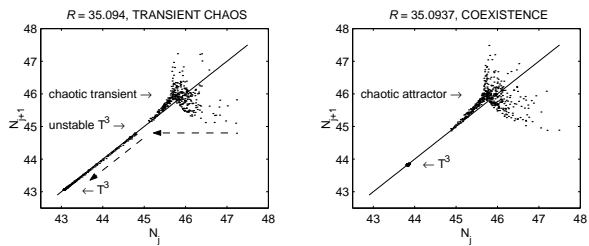}
\caption[]{Return map of the maxima of the variable $Z_0$
satisfying to be larger than their adjacent maxima (see text). The
left panel shows a regime of transient chaos; after a transient
the trajectory decays to the stable $\mathbb{T}^3$ (following the
dotted arrows). As $R$ is decreased the stable and the unstable
$\mathbb{T}^3$ get closer. Beyond some point ($R<R_{bc}$)  orbits
inside the chaotic set do not escape, which means that the chaotic
set has become an attractor. The right panel shows the attractors
occurring for $R$ into the coexistence interval.} \label{map}
\end{figure}

In our case we may try to reduce the dynamics by means of a return
map. The fast dynamics concerning the $k=1$ mode is approximately
filtered when considering the $k=0$ mode. Then, the $\mathbb{T}^3$
attractor is seen in the $k=0$ framework as a $\mathbb{T}^2$ plus
some small residual oscillating component. A return map of the variable
$Z_0$ would reduce the dimension of the attractors by one, and
giving as a result an (approximately) one dimensional curve for 
the $\mathbb{T}^3$ attractor. 
We must take an additional
return map to reduce the $\mathbb{T}^3$ to a fixed point, 
and the chaotic set of dimension (about) four to a line.
Considering the set of maxima of $Z_0(t)$, $\{M_n\}$, we took the
subset of maxima whose neighboring maxima were smaller:
$\{N_j\}=\{M_n> M_{n \pm 1}\}$,
what effectively amounts to consider the maxima of the low frequency
oscillations.
The results for two values of the parameter $R$ at both sides of the crisis
($R=R_{bc}$) are shown in
Fig.~\ref{map}. The chaotic attractor exhibits a rough
$\Lambda$-shaped structure as occurs with the Lorenz map.
Probably, the existence of the residual fast component makes the
attractor to deviate significantly from one-dimensionality. In the
light of the paper by Yorke and Yorke~\cite{yorkey}, who studied
the transition to sustained chaotic behavior in the Lorenz model
with the Lorenz map, it is found that our results are consistent
with a boundary crisis mediated by an unstable three-torus.

\section{Route to chaos: theoretical analysis}
\label{chaosroute}

Condensing all the information obtained from numerical experiments
in the previous sections, we suggest the route to high-dimensional
chaos represented in Fig.~\ref{3dscheme} (where it is to be
understood that the customary cross
section through the fast rotating wave is applied). The high
dimension of our attractor makes somewhat convoluted a geometric
visualization. As we explained above, we postulate a chaotic
attractor whose structure may be simplified in terms of a
4-dimensional branched manifold, and therefore the Poincar\'e
section reduces the attractor to a 3-dimensional branched
manifold. Figure~\ref{3dscheme} represents a projection onto
$\mathbb{R}^3$, hence some (apparently) forbidden intersections
between trajectories appear because of the branching. As occurs
with the Lorenz attractor when it is projected on $\mathbb{R}^2$
(say $x-z$), the intersections between trajectories coming from
different lobes, and also of them with the $z$ axis (that belongs
to the stable manifold of the origin) are unavoidable. Recall that
it is the moment of the jump when the extra dimension is needed,
and this is provided by the definition of a branched manifold.

\begin{widetext}
\begin{center}
\begin{figure}
\includegraphics[width=0.9\columnwidth]{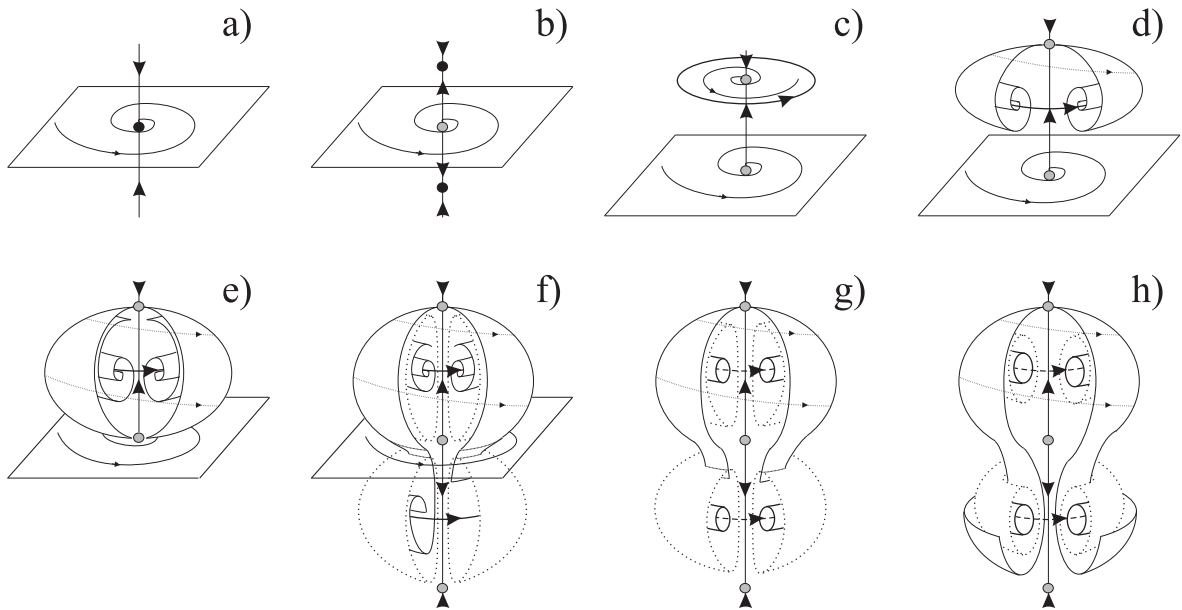}
\caption[]{Three-dimensional representation of the proposed
heteroclinic route to create the high-dimensional chaotic
attractor. Black and gray points correspond to stable and unstable
fixed points (cycles in the global phase space), respectively.
This representation is obtained through a Poincar\'e cross section
of the real attractors of the system, that eliminates the (fast) 
oscillation along the ring.
}
\label{3dscheme}
\end{figure}
\end{center}
\end{widetext}

Summing up our previous numerical findings, we have 
for descending values of $R$: the centered PRW (a)
becomes unstable through a pitchfork bifurcation (a$\rightarrow$b)
and two symmetry-related PRWs appear (b). At a supercritical Hopf
bifurcation (b$\rightarrow$c) the $\mathbb{T}^2$ appears. When $R$
is slightly decreased the 2D-torus becomes focus-type (d); the leading
Lyapunov exponent becomes degenerate as may be seen in
Fig.~\ref{lyap_tran}. Hence the unstable manifold of the
asymmetric PRW forms a `whirlpool' \cite{shilnikov95} when approaching the
$\mathbb{T}^2$. At $R_{het}$ a double heteroclinic connection
between the asymmetric PRWs and the symmetric one occurs (e)
(only the upper half of the connection is shown for clarity). At
this point the chaotic set is created, that includes a dense set
of unstable 3D-tori. In (f) the two simplest 3D-tori are
represented with dotted lines, because of the heteroclinic birth
one of the frequencies of these tori is very small. Note that the
plot shows that the unstable manifold of one of the asymmetric
PRWs intersects the stable manifold of the symmetric PRW which in
principle violates the theorem of existence and uniqueness. This
occurs, as we said above, because we are projecting the Poincar\'e
section onto $\mathbb{R}^3$ (one could imagine this as our
particular {\it Flatland\/}\footnote{Making an analogy to the imagined 
world in Edwin Abbott's book.}. Twin
secondary Hopf bifurcations (f$\rightarrow$g) render unstable the
2D-tori and give rise to two stable $\mathbb{T}^3$ (g). When $R$
is further decreased the asymmetric PRWs are not connected
by their unstable manifolds  
to the $\mathbb{T}^3$ (h), and the chaotic set becomes attracting.

\begin{figure}
\includegraphics[width=0.9\columnwidth]{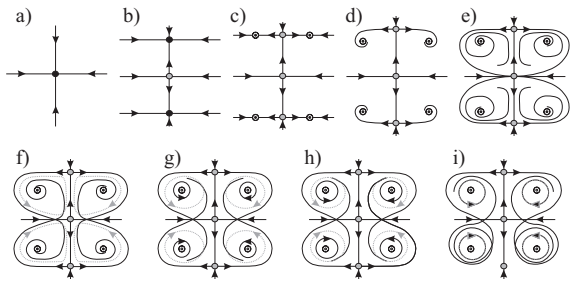}
\caption[]{Two-dimensional representation of the proposed
heteroclinic route to create the high-dimensional chaotic
attractor. It represents a vertical cut of Fig.~\ref{3dscheme}.
$\odot$ (resp.~$\circledcirc$) symbols
represent stable (resp.~unstable) periodic orbits 
(2D-tori in the full phase space).}
\label{2dscheme}
\end{figure}

A two-dimensional cut of the schematic shown in
Fig.~\ref{3dscheme} is depicted in Fig.~\ref{2dscheme}. Only some of the 
subplots of both figures are one-to-one related. 
Specifically, Fig.~\ref{2dscheme}(h) corresponds to $R=R_{bc}$,
whereas Fig.~\ref{2dscheme}(i) corresponds to $R=R_{sn}$.

It is to be stressed that the existence of $\mathbb{T}^3$ attractors
is not a fundamental part of the transition to high-dimensional
chaos. Just focus type $\mathbb{T}^2$ are needed 
such that the unstable manifolds of the asymmetric PRWs form whirlpools
\cite{shilnikov95}. 
In this way, regarding the chaotic attractor, no fundamental change
occurred if the unstable $\mathbb{T}^3$ shrank to collide with the
stable $\mathbb{T}^2$ in a subcritical Neimark-Sacker bifurcation. This
picture would be more similar to the transition in the Lorenz
system where $C_{\pm}$ become unstable through a subcritical Hopf
bifurcation.

\section{Further remarks}
\label{further}

In this section we want to address some subtle aspects concerning
the transition to chaos shown in this paper.

The first point that must be noted is that Fig.~\ref{3dscheme} is
representing a map (instead of a continuous dynamical system)
because a Poincar\'e section of the fast dynamics involving the
spatial mode is assumed. 
However the fast spatial frequency is somehow `orthogonal'
along the transition and therefore phase space may be understood
as a direct product of this frequency with the transition shown in
Fig.~\ref{3dscheme} that can be assumed not to differ too much
from a continuous system.
The fast spatial rotating wave is approximately conserved 
along the transition (cf. Sec.~\ref{quasip}), and the 
weak interaction of 
this frequency stems from the rotational symmetry of the system and
manifests through
an additional
vanishing Lyapunov exponent in the high-dimensional chaotic
region. 

Let us consider further which is the meaning of the Lyapunov
spectrum of the chaotic region. According to the route described
by Figs.~\ref{3dscheme} and \ref{2dscheme} an infinite number of 
unstable $\mathbb{T}^3$ are created at the heteroclinic explosion 
(recall that only the two simplest ones are shown). Hence, we 
could expect to have a chaotic attractor with three vanishing 
Lyapunov exponents. Why only two are observed? (cf.~region CRW in 
Fig.~\ref{lyap_tran}).

One could be tempted to think that that due to the
non-hyperbolicity of quasiperiodic behavior, a finite ratio of the
3D-tori set corresponds to locked tori (probably with large
denominators), contributing to shift the vanishing Lyapunov
exponents from zero.
However we believe that the absence of a third vanishing Lyapunov
exponent is due to the following. It must be noted that the
heteroclinic connection between the symmetric and the asymmetric
PRWs is structurally unstable. 
Nonetheless, we have numerically observed (see
Fig.~\ref{explosion}) that the connection (approximately)
persists. Also, the axial symmetry of Fig.~\ref{3dscheme} has not
a theoretical justification.
In our view the situation is similar to the study of
codimension-$m$ ($m>1$) bifurcations,
where an increase of the highest order $n<n_{max}$, of the
terms constituting the normal form (or $n$-jet) or adding a non
symmetric term, makes substructures in the bifurcation sets emanating
from the codimension-$m$ point to appear. For example, the
effect of including non-axisymmetric terms to the normal form of the
saddle-node Hopf (codimension-two) bifurcation has been studied by
Kirk~\cite{kirk91}.

Hence we expect that, a perturbation on the mechanism shown in
Fig.~\ref{3dscheme} will break the symmetry that allows such a 
``simple'' picture. Inspired on
previous works~\cite{kirk91,gaspard} dealing with the effect of
non-symmetric terms on codimension-two bifurcations, we postulate
that homoclinic connections would replace heteroclinic connections.
A double (`figure-eight') homoclinic of the symmetric PRW as well
as homoclinic connections of the asymmetric PRWs would occur.
Symmetric and asymmetric PRWs are both saddle-foci so we  
computed their saddle indices
finding that a (Shilnikov) chaotic set
should arise\footnote{For a
saddle-focus with three eigenvalues $\lambda_s= \rho \pm i \omega$
($\rho<0$), $\lambda_u>0$ (as the symmetric PRW), the saddle index
is defined to be $\delta=- \rho / \lambda_u$, and chaos will occur
for $\delta<1$. Furthermore, no stable periodic orbits exist in the
neighborhood of homoclinicity for $\delta<1/2$ \cite{GlendinningPLA}.
In our case, 
symmetric and asymmetric PRWs are both
fixed points of the 8-dimensional map (obtained via Poincar\'e section),
and, thus, they can be stabilized through the use of a Newton method.
As the sectioning is done to eliminate the fast rotating wave, we are legitimate to
consider a continuous 8-dimensional system. The eigenvalues of the fixed points (PRWs in the full space)
satisfy: $Re(\lambda_i)=(1/T)\log|\mu_i|$ ($\{\mu_i\}$ are the eigenvalues of the map).
We take the leading eigenvalues to compute the saddle indices. 
Thus for $R=35.1$, 
we get a saddle index $\delta= 0.43...$ 
for the symmetric PRW,
and under time reversal $\delta=1.32...$
for the asymmetric PRW. These values of $\delta$ imply
the emergence of a chaotic set and absence of stable periodic orbits 
close to the symmetric PRW; and one repeller (due to time-reversal) 
periodic orbit colliding with the asymmetric PRW.}.

The existence of homoclinic chaos (with the only novel 
feature of having a superimposed fast spatial wave) could be considered
as uninteresting and well known (see e.g.~\cite{arneodo}). 
But it is to be emphasized
that as the exact mechanism is very related to what shown
in Fig.~\ref{3dscheme}, the first negative Lyapunov exponent is
very close to zero which makes the information dimension to be
larger than four (or larger than three if the spatial oscillation
is considered extra and/or trivial).

The next question is how the two simplest unstable 3D-tori appear if the
route is not exactly as appears in Fig.~\ref{3dscheme}. A
possibility is that they appear though a saddle-node bifurcation
between two unstable 2-tori (see footnote 6 for its
plausible origin), but we have no way to find out this.

It is well known, \cite{Guckenheimer,Kuznetsov,gaspard} that global bifurcations
and complex behavior
have in many cases a local origin, namely a high codimension point.
at which the loci of several bifurcations meet in parameter space.
In particular, a Gavrilov-Guckenheimer (or saddle node-Hopf point) yields,
among other, quasiperiodic dynamics. The reflection symmetry of the
model studied in this work would imply the possibility of a pitchfork-Hopf
interaction in this case. Although the unfolding for this codimension-$2$
point has not been fully characterized, it could be a clear candidate
as an organizing center of part of the complex behavior discussed in
this work. However, the location of such hypothetical critical point is
rendered more difficult by the fact that, due to the presence of a spatial 
frequency in all the relevant parameter region, the hypothetical pitchfork-Hopf
point would occur in a Poincar\'e cross section of the system, making all
the analysis much more involved. In particular, we have tried this avenue
of research without success, and, thus,
we have not been able to find a coalescence of
the loci of the pitchfork and Hopf bifurcations.

\section{Conclusions} \label{discussion}

Direct transitions from quasiperiodicity to chaos has been observed experimentally
(e.g. in \cite{dubois}). In the past this has been interpreted as an effect of noise or 
lack of good control of the system. Our results indicate that high-dimensional
chaos may be reached directly without the need of noise, but only thanks to a 
particular type of global bifurcation.

In this paper we have studied by numerical and theoretical
arguments the transition to high-dimensional chaos, in a system of
three coupled Lorenz oscillators. The transition from a periodic
rotating wave to a chaotic rotating wave has been investigated.
The structure of the global bifurcations between cycles,
underlying the creation of the chaotic set, is such that a chaotic
attractor with dimension $d\approx4$ emerges. The transition is
not mediated by low-dimensional chaos.  Also it must be noted that
even  if the fast rotating wave that is present all along the
transition is omitted, we still have the creation of an
attractor with dimension $D_1 \gtrsim 3$. As occurs with the
Lorenz system, the existence of reflection ($\mathbf{Z}_2$)
symmetry seems to play a fundamental role.

The high-dimensionality of the chaotic attractor is not associated to
hyperchaos. Far from it, there is only one positive Lyapunov
exponents but high-dimensionality is possible due to the existence
of two vanishing and one slightly negative LEs. Hence, according
to the Kaplan-Yorke conjecture the information dimension is above
four. We have also measured
the correlation dimension obtaining a value very close
to four. The degeneracy of the null LE make us think that a set of
unstable tori is embedded in the attractor.

We have focused on giving a geometric view of the bifurcations
occurring in the 9-dimensional phase space of the system. Although
the precise sequence of bifurcations is probably resistant to
analysis\footnote{Quasi-attractors (as those appearing through 
homoclinicity to a saddle-focus point) exhibit infinitely many bifurcations of 
various types and cannot be described completely 
~\cite{gonchenko}.}, we have been able to give a geometric view of the
transitions that explains the emergence of the chaotic set,
through a `heteroclinic explosion', and its conversion in
attractor. This step occurs through a boundary crisis when the
chaotic attractor collides with its basin boundary formed by two
unstable 3D-tori. In consequence a power law for the mean length
of the chaotic transients is observed.

This work was supported by MICINN (Spain) and FEDER (EU) under
Grants No. BFM2003-07749-C05-03 and FIS2006-12253-C06-04 (DP),
and FIS2007-60327 (FISICOS) (MAM).


\newcommand{\noopsort}[1]{} \newcommand{\printfirst}[2]{#1}
  \newcommand{\singleletter}[1]{#1} \newcommand{\switchargs}[2]{#2#1}

\end{document}